\documentclass[preprint,showpacs,showkeys,preprintnumbers,amsmath,amssymb,,nofootinbib]{revtex4}
\usepackage{amssymb}
\usepackage{amsfonts}
\usepackage{amsmath}

\usepackage{graphicx}
\usepackage{dcolumn}
\usepackage{bm}

\newcommand{\be}{\begin{equation}}
\newcommand{\bea}{\begin{eqnarray}}
\newcommand{\eea}{\end{eqnarray}}
\newcommand{\ba}{\begin{array}}
\newcommand{\ea}{\end{array}}
\newcommand{\ee}{\end{equation}}

\def\bse{\begin{subequations}}
\def\ese{\end{subequations}}

\begin{document}

\title{Disk relations for tree amplitudes in minimal coupling theory of gauge field and gravity}

\author{Yi-Xin Chen}
\thanks{Corresponding author} \email{yxchen@zimp.zju.edu.cn}
\author{Yi-Jian Du}
\email{yjdu@zju.edu.cn}
\author{Qian Ma}
\email{mathons@gmail.com}

\affiliation{Zhejiang Institute of Modern Physics, Zhejiang
University,
 Hangzhou 310027, P. R. China}

\begin{abstract}KLT relations on $S_2$ factorize closed string amplitudes into product of open string tree amplitudes.
The field theory limits of KLT factorization relations hold in
minimal coupling theory of gauge field and gravity. In this paper,
we consider the field theory limits of relations on $D_2$. Though
the relations on $D_2$ and KLT factorization relations hold on
worldsheets with different topologies, we find the field theory
limits of $D_2$ relations also hold in minimal coupling theory of
gauge field and gravity. We use the $D_2$ relations to give three-
and four-point tree amplitudes where gluons are minimally coupled to
gravitons. We also give a discussion on general tree amplitudes for
minimal coupling of gauge field and gravity. In general, any tree
amplitude with $M$ gravitons in addition to $N$ gluons can be given
by pure-gluon tree amplitudes with $N+2M$ legs.
\end{abstract}
\date{\today}
 %  \\
 \maketitle
\section{Introduction}
Superstring theories are theories containing both gravitational and
gauge interactions\cite{1, 2}. They offer a possible way to unify
the gauge field and gravity. In string theory, gravitons and gauge
particles can correspond to massless states of closed and open
strings. Then the relations between closed and open strings imply
the relations between gravity and gauge field.

There are many relationships between gauge field and gravity in
string theory such as AdS/CFT\cite{3} correspondence and the
perturbative relations for amplitudes on sphere($S_2$)\cite{4},
disk($D_2$)\cite{5, 6, 7, 8, 9, 10} and real projective
plane($RP_2$)\cite{5}. The perturbative relations in $S_2$ case
named KLT relation\cite{4} factorize the amplitudes on $S_2$ into
products of two amplitudes for open strings corresponding to the
left- and the right-moving sectors of closed strings. However, the
amplitudes on $D_2$ cannot be factorized into two sectors, because
the boundary of $D_2$ connect the two sectors into a single one. We
should use the new relations instead of KLT factorization relations
on $D_2$. KLT factorization relations and $D_2$ relations hold on
worldsheets with different topologies.

In field theory limit, KLT factorization relations allow one to
obtain gravity amplitudes from gauge theory ones\cite{11, 12, 13,
14, 15, 16}. They can also be used in many theories of
gravity-matter couplings\cite{17, 18}. An important application is
that the low energy limits of KLT relations can be used to calculate
the tree amplitudes for gluons minimally coupled to gravitons. In
this case, using the relations, the amplitudes with gluons and
gravitons can be factorized into products of amplitudes in left- and
right-moving sectors. Amplitudes in one sector are pure gauge
partial amplitudes while those in the other sector are partial
amplitudes with gluons and scalars. These relations for
gauge-gravity minimal coupling are based on the structure of
heterotic strings\cite{1, 2, 19, 20, 21}. In fact, in heterotic
string theories, gauge degrees of freedom are taken by Lorentz
singlets in one sector of closed strings.

There is another way to incorporate gauge degrees of freedom into
string theory. In theories containing open strings such as Type I
theory, one can add Chan-Paton factors\cite{1, 2, 22, 23} to the
ends of open strings. The interactions between closed and open
strings at tree-level are on $D_2$. In the field theory limits of
this case, the tree amplitudes for gauge-gravity interaction and
those for pure gauge field are connected via $D_2$ relations. Then a
question arises: In what kind of theory for gauge-gravity
interactions do the field limits of $D_2$ relations hold? Or do the
field theory alimits of $D_2$ relations also hold in minimal
coupling theory of gauge and gravity?

In this paper, we study the amplitudes in minimal coupling theory of
gauge field and gravity. We find though the KLT factorization
relations and the $D_2$ relations hold on worldsheets with different
topologies in string theory, the field theory limits of $D_2$
relations also hold in minimal coupling theory of gauge field and
gravity. $D_2$ relations in minimal coupling theory of gauge field
and gravity are based on the disk structure. They give a new
understanding on gauge-gravity interaction.

The structure of this paper is as follows. In Section \ref{KLT vs
disk}, we give an introduction to KLT relations and $D_2$ relations.
We also give the low energy limits of $D_2$ relations for amplitudes
with three and four legs. In Section \ref{3,4point}, we use the low
energy limits of $D_2$ relations to give the tree amplitudes for
gauge-gravity interaction. We find the amplitudes given by $D_2$
relations are same with those given by KLT factorization relations.
Then the $D_2$ relations give the amplitudes in minimal coupling
theory of gauge field and gravity. In Section \ref{general
discussions}, we consider the $D_2$ relations for general tree
amplitudes where gluons are minimally coupled to gravitons. We first
study the mixed amplitudes where all the legs take positive helicity
or only one leg take negative helicity. In these cases, the $D_2$
relations hold trivially. Then we study the tree amplitudes with one
and two gravitons in addition to $N$ gluons where $N-2$ gluons as
well as all the gravitons take positive helicity and two gluons take
negative helicity. We will show the $D_2$ relations also hold in
this case. The discussions can be extended to tree amplitudes where
$N$ gluons minimally coupled to $M$ gravitons with arbitrary
helicity configurations. In general, any tree amplitudes for
gauge-gravity minimal coupling can be expressed by partial tree
amplitudes with $N+2M$ gluons via $D_2$ relations. Our conclusions
are given in Section \ref{conclusion}. Some useful properties of
spinor helicity formalism are given in Appendix \ref{spinor
helicity}.

\section{KLT relations versus $D_2$ relations}\label{KLT vs disk}
KLT relations\cite{4} in string theory are the relations between
amplitudes for closed strings on $S_2$ and open string tree
amplitudes. KLT relations factorize amplitudes for closed strings on
$S_2$ into products of two open string tree amplitudes corresponding
to the left- and right-moving sectors except for a phase
factor\footnote{In this paper, we use $\sim$ to omit a proportional
factor which does not affect our discussion.}
\begin{equation}\label{KLT relation}
\mathcal {M}_{S_2}^{(N)}\sim\kappa^{N-2} \sum\limits_{P, P'}\mathcal
{A}^{(N)}(P)\bar{\mathcal {A}}^{(N)}(P')e^{i\pi F(P, P')},
\end{equation}
Where $\mathcal {M}_{S_2}^{(N)}$ is $N$-point amplitude on $S_2$
while $\mathcal {A}^{(N)}$ and $\bar{\mathcal {A}}^{(N)}$ are the
partial tree amplitudes for open strings corresponding to the left-
and right-moving sectors. The phase factor only depends on the
permutations $P$ and $P'$ of the legs in left- and right-moving
sectors. The terms in KLT relations can be reduced by contour
deformations. In the reduced form, the phase factors become sine
functions. After taking the field theory limit $\alpha'\rightarrow
0$, the KLT relations for three- and four-point amplitudes are given
as
\begin{subequations}
\begin{align}
\label{KLT 3point}\mathcal {M}(1, 2, 3)&\sim\kappa\mathcal{A}(1, 2,
3)\bar{\mathcal{A}}(1, 2, 3),
\\\label{KLT 4point}\mathcal {M}(1, 2, 3, 4)&\sim\kappa^2(-i)s_{12}\mathcal {A}(1, 2, 3, 4)\bar{\mathcal
{A}}(1, 2, 4, 3).
\end{align}
\end{subequations}

In field theory limits, KLT relations factorize the pure-graviton
tree amplitudes into products of tree amplitudes for gluons
corresponding to left- and right-moving sectors. The tree amplitudes
where gravitons are minimally coupled to gluons can also be
factorized by KLT relations. In this situation, the two sectors of a
graviton with helicity $\pm2$ correspond to two gluons with helicity
$\pm1$, while the two sectors of a gluon with helicity $\pm1$
correspond to one gluon with helicity $\pm1$ and one scalar
particle. The gauge degrees of freedom are taken by the scalar field
in one sector. With these correspondences, KLT relations express the
amplitudes for $N$ gluons and $M$ gravitons by products of
amplitudes in left- and right-moving sectors. The amplitudes in the
left-moving sector are pure-gluon partial tree amplitudes with $N+M$
legs while the amplitudes in the right-moving sector are partial
tree amplitudes for $M$ gluons and $N$ scalar particles. Using the
Feynman rules given in \cite{17}, one can calculate the partial
amplitudes in the left- and right-moving sectors, then the
amplitudes where gluons are minimally coupled with gravitons can be
given by KLT relations.

The KLT factorization relations in string theory hold on $S_2$. But
they do not hold on $D_2$. In $D_2$ case, the left- and right-moving
sectors of closed strings are connected into a single one\cite{5}.
The relations between amplitudes for $N$ open strings in addition to
$M$ closed strings on $D_2$ and open string tree amplitudes are
given as
\begin{equation}\label{Disk relation}
 \mathcal{M}_{D_2}^{(N,
M)}\sim g^{N-2}\kappa^{M}
\sum\limits_P\mathcal{A}^{(N,2M)}(P)e^{i\pi\Theta'(P)}.
\end{equation}
In the relation \eqref{Disk relation}, we do not introduce the
Chan-Paton degrees of freedom.
 On $D_2$, one can add Chan-Paton
factor to the ends of open strings to incorporate gauge degree of
freedom. Then the amplitudes on $D_2$ can be given by color
decomposed form
\begin{equation}\label{color decomposition}
\begin{split}
&\mathcal {M}(1_o^{a_1},...,N_o^{a_N}, (N+1)_c, ..., (N+M)_c)
\\=&\sum_\sigma
Tr\left(T^{a_{\sigma}(1)}...T^{a_{\sigma}(N)}\right)\mathcal
{A}(\sigma(1_o),...,\sigma(N_o),(N+1)_c,...,(N+M)_c),
\end{split}
\end{equation}
where $i_o^a(i=1,...,N)$ denote open string legs with Chan-Paton
degrees of freedoms and $j_c(j=N+1,...,N+M)$ denote closed string
legs. $\sigma$ runs over the set of non cyclic permutations of the
open strings. Then $D_2$ relations give the partial amplitudes
$\mathcal {A}(\sigma(1_o),...,\sigma(N_o),(N+1)_c,...,(N+M)_c)$ for
a given permutation of open strings by pure open string amplitudes:
\begin{equation}\label{partial disk relations}
\mathcal
{A}(\sigma(1_o),...,\sigma(N_o),(N+1)_c,...,(N+M)_c)=\sum\limits_P
e^{i\pi\Theta'(P'')}\mathcal {A}^{(N, 2M)}(P'').
\end{equation}
where $P''$ are all the permutations of the $N+2M$ external legs
which preserve the relative positions of the open strings
$1_o,...,N_o$. This expression implies that for a given permutation
of the open strings on the boundary of $D_2$, any closed string can
split into two open strings inserted on the boundary of $D_2$. Using
contour deformations, the relations \eqref{partial disk relations}
can also be reduced\cite{6}. In the reduced form of the relations,
the phase factors become sine functions.

In field theory limits, $D_2$ relations give tree amplitudes with
$N$ gluons and $M$ gravitons by pure-gluon amplitudes with $N+2M$
legs. They are different from the KLT factorization relations which
factorize amplitudes with $N+M$ legs into products of two amplitudes
with $N+M$ legs. For $M=0$, $D_2$ relations trivially give the
pure-gluon amplitudes. Then we do not need to consider $M=0$ case.
Since the generators of the gauge group satisfy $Tr(T^a)=0$, we also
do not need to consider $M=1$ case. After taking the field theory
limit $\alpha'\rightarrow 0$, the $D_2$ relations for two-gluon
one-graviton tree amplitude $\mathcal{A}(1_g, 2_g, 3_h)$,
three-gluon one-graviton tree amplitude $\mathcal{A}(1_g, 2_g, 3_g,
4_h)$ and two-gluon two-graviton tree amplitude $\mathcal{A}(1_g,
2_g, 3_h, 4_h)$ are given as\footnote{In this paper, we use $i_g$
and $j_h$ to denote gluons and gravitons respectively. We do not
consider the amplitudes where all the external legs are gravitons.
We also do not consider the amplitudes where graviton exchanges
between gluons, which contribute to higher order process. }
\begin{subequations}\label{3point and 4point relations}
\begin{align}
\label{2g1h}\mathcal{A}(1_g, 2_g, 3_h)&\sim\kappa s_{12}\mathcal
{A}(1_g, 2_g, 3_g, 4_g),
\\\label{3g1h}\mathcal{A}(1_g, 2_g, 3_g, 4_h)&\sim g\kappa s_{13}\mathcal {A}(1_g, 5_g, 2_g, 4_g,
3_g),
\\\label{2g2h}\mathcal{A}(1_g, 2_g, 3_h, 4_h)&\sim \kappa^2[ s_{12}^2
\mathcal {A}(1_g, 6_g, 3_g, 5_g, 4_g, 2_g) -s_{12}s_{13}\mathcal
{A}(1_g, 3_g, 5_g, 4_g, 2_g, 6_g)],
\end{align}
\end{subequations}
where  $s_{ij}=2k_i\cdot k_j$, $3_g$ and $4_g$ have momentum
$\frac{1}{2}k_3$ while $5_g$ and $6_g$ have momentum
$\frac{1}{2}k_4$. The total amplitude is derived by substitute the
relations \eqref{3point and 4point relations} into Eq. \eqref{color
decomposition}.

So far, we have seen, though KLT and $D_2$ relations hold on
worldsheets with different topologies in string theory. They can
both give the amplitudes for gauge-gravity coupling. KLT relations
can given the amplitudes for gauge-gravity minimal coupling. Then we
should consider the question: Can the field theory limits of the two
different relations in string theory hold in a same theory for
gauge-gravity coupling? In the next two sections, we will show the
$D_2$ relations also hold in minimal coupling theory of gauge field
and gravity.

\section{$D_2$ relations for three- and four-point tree amplitudes}\label{3,4point}
In this section, we use the $D_2$ relations \eqref{3point and 4point
relations} to give the three- and four-point tree amplitudes for
gauge-gravity coupling. These results are same with those given by
using KLT relations\cite{17}. Thus $D_2$ relations also hold in
minimal coupling theory of gauge field and gravity.

\subsection{Three-point tree amplitude}
The only nontrivial three-point partial amplitude needed is
$\mathcal {A}(1_g, 2_g, 3_h)$. To give this amplitude, we need to
calculate the amplitudes $\mathcal {A}(1_g, 2_g, 3_g, 4_g)$ in which
$3_g$ and $4_g$ correspond to the left- and right-moving sectors of
the graviton $3_h$. Using the color-ordered Feynman rules in
\cite{24} and replacing the momenta $k_3$ and $k_4$ by
$\frac{1}{2}k_3$, we can get the amplitude $\mathcal {A}(1_g, 2_g,
3_g, 4_g)$. After substituting it into the relation \eqref{2g1h},
the three-point amplitude $\mathcal {A}(1_g, 2_g, 3_h)$ is given
\begin{equation}
\mathcal {A}(1_g, 2_g, 3_h)
\sim2[-\epsilon_1\cdot\epsilon_2\epsilon_3^{\rho\sigma}k_{1\rho}k_{2\sigma}+\epsilon_1\cdot
k_2\epsilon_{3\rho\sigma}k_{1\rho}\epsilon_{2\sigma}+\epsilon_2\cdot
k_1\epsilon_{3}^{\rho\sigma}\epsilon_{1\rho}k_{2\sigma}],
\end{equation}
where the physical conditions $\epsilon_1\cdot k_1=\epsilon_2\cdot
k_2=0$,
$\epsilon_{3\rho\sigma}k_3^{\rho}=\epsilon_{3\rho\sigma}k_3^{\sigma}=0$,
momentum conservation $k_1^{\mu}+k_2^{\mu}+k_3^{\mu}=0$ and the
traceless condition of the polarization tensor of graviton
$\epsilon_{3\rho}^{\rho}=0$ have been used. This is same with that
given by KLT relation\eqref{KLT 3point}. Thus, $D_2$ relation can
give the three-point tree amplitude where gluons are minimally
coupled to graviton.
\subsection{Four-point tree amplitudes}
In this subsection, we study the four-point amplitudes
$\mathcal{A}(1_g, 2_g, 3_g, 4_h)$ and $\mathcal{A}(1_g, 2_g, 3_h,
4_h)$. We use the spinor helicity formalism\cite{24, 25, 26} to
consider the four-point amplitudes. Useful properties of spinor
helicity formalism are listed in Appendix \ref{spinor helicity}.

The two gluons corresponding to a graviton with helicity $\pm2$ take
helicity $\pm1$. Then the four-point tree amplitudes with all legs
of positive helicity can be given by pure-gluon amplitudes with all
legs of positive helicity via the relations \eqref{3g1h} and
\eqref{2g2h}. Because the pure-gluon partial amplitudes in which all
gluons take the same helicity vanish, we have
\begin{equation}
\mathcal{A}(1_g^+, 2_g^+, 3_g^+, 4_h^+)=\mathcal{A}(1_g^+, 2_g^+,
3_h^+, 4_h^+)=0.
\end{equation}

The tree amplitudes with one gluon of negative helicity and other
legs of positive helicity can be given by pure-gluon tree amplitudes
where only one leg take negative helicity and other legs take
positive helicity. Since the pure-gluon tree amplitudes with one leg
of negative helicity and other legs of positive helicity vanish, we
have
\begin{equation}
\mathcal{A}(1_g^-, 2_g^+, 3_g^+, 4_h^+)=\mathcal{A}(1_g^-, 2_g^+,
3_h^+, 4_h^+)=0.
\end{equation}
The tree amplitudes with one graviton of negative helicity and other
legs of positive helicity can be given by pure-gluon MHV\cite{27}
amplitudes. In this pure-gluon MHV amplitude, the two gluons
corresponding to the negative helicity graviton take negative
helicity. Using the relation \eqref{3g1h} and the expression of MHV
amplitude for gluons \eqref{MHV gluon amplitude}, the amplitude
$\mathcal{A}(1_g^+, 2_g^+, 3_g^+, 4_h^-)$ is given
\begin{equation}
\mathcal{A}(1_g^+, 2_g^+, 3_g^+, 4_h^-)\sim g^2\kappa\mathcal
{A}(1_g^+, 5_g^-, 2_g^+, 4_g^-, 3_g^+)=g \kappa
s_{13}i\frac{\langle54\rangle^4}{\langle15\rangle\langle52\rangle\langle24\rangle\langle43\rangle\langle31\rangle}=0,
\end{equation}
where we have use $k_5=k_4$. Similarly, $\mathcal{A}(1_g^+, 2_g^+,
3_h^-, 4_h^+)=0$.

Now we consider the MHV amplitudes where two legs take negative
helicity and others take positive helicity. There are three
independent amplitudes $\mathcal{A}(1_g^-, 2_g^-, 3_g^+, 4_h^+)$,
$\mathcal{A}(1_g^-, 2_g^-, 3_h^+, 4_h^+)$ and $\mathcal{A}(1_g^-,
2_g^+, 3_h^-, 4_h^+)$. With the $D_2$ relations \eqref{3g1h} and
\eqref{2g2h}, the first two amplitudes can be expressed by five- and
six-point pure-gluon MHV amplitudes respectively. After using some
properties of the spinor helicity formalism and the fact that the
two gluons corresponding to one graviton take half of the momentum
of the graviton, we get the first two amplitudes
\begin{equation}
\begin{split}
\mathcal{A}(1_g^-, 2_g^-, 3_g^+, 4_h^+)&\sim g\kappa s_{13}\mathcal
{A}(1_g^-, 5_g^+, 2_g^-, 4_g^+, 3_g^+)
\\&=g\kappa s_{13}i\frac{\langle12\rangle^4}{\langle15\rangle\langle52\rangle\langle24\rangle\langle43\rangle\langle31\rangle}
\\&\sim
g\kappa\sqrt{2}\frac{\langle12\rangle^4[12]}{\langle14\rangle\langle24\rangle\langle34\rangle^2}.
\end{split}
\end{equation}
\begin{equation}
\begin{split}
\mathcal{A}(1_g^-, 2_g^-, 3_h^+, 4_h^+)&\sim \kappa^2[ s_{12}^2
\mathcal {A}(1_g^-, 6_g^+, 3_g^+, 5_g^+, 4_g^+, 2_g^-)
-s_{12}s_{13}\mathcal {A}(1_g^-, 3_g^+, 5_g^+, 4_g^+, 2_g^-, 6_g^+)]
\\&=\kappa^2\left[{s_{12}^2i\frac{\langle12\rangle^4}{\langle16\rangle\langle63\rangle\langle35\rangle\langle54\rangle\langle42\rangle\langle21\rangle}}
-s_{12}s_{13}\frac{\langle12\rangle^4}{\langle13\rangle\langle35\rangle\langle54\rangle\langle42\rangle\langle26\rangle\langle61\rangle}\right]
\\&=0.
\end{split}
\end{equation}
To calculate $\mathcal{A}(1_g^-, 2_g^+, 3_h^-, 4_h^+)$, we need
six-point non-MHV tree amplitudes for gluons $\mathcal {A}(1_g^-,
6_g^+, 3_g^-, 5_g^+, 4_g^-, 2_g^+)$ and $\mathcal {A}(1_g^-, 3_g^-,
5_g^+, 4_g^-, 2_g^+, 6_g^+)$. Using the tree amplitude with six
gluons given in\cite{25} and substituting  $k_3$, $k_4$ and $k_5$,
$k_6$ by $\frac{k_3}{2}$ and $\frac{k_4}{2}$ correspondingly, we get
\begin{equation}\label{6gNMHV}
\begin{split}
\mathcal {A}(1_g^-, 6_g^+, 3_g^-, 5_g^+, 4_g^-,
2_g^+)&=-16i\frac{[24]^4\langle13\rangle^2\langle23\rangle^2}{s_{12}^3s_{23}^2},
\\\mathcal {A}(1_g^-, 3_g^-,
5_g^+, 4_g^-, 2_g^+,
6_g^+)&=16i\frac{[24]^4\langle13\rangle^2\langle23\rangle^2}{s_{12}s_{13}^2s_{23}^2}.
\end{split}
\end{equation}
Then we substitute the amplitudes\eqref{6gNMHV} into the
relation\eqref{2g2h}. The amplitude $\mathcal{A}(1_g^-, 2_g^+,
3_h^-, 4_h^+)$ is given
\begin{equation}
\begin{split}
\mathcal{A}(1_g^-, 2_g^+, 3_h^-, 4_h^+)&\sim \kappa^2[ s_{12}^2
\mathcal {A}(1_g^-, 6_g^+, 3_g^-, 5_g^+, 4_g^-, 2_g^+)
-s_{12}s_{13}\mathcal {A}(1_g^-, 3_g^-, 5_g^+, 4_g^-, 2_g^+, 6_g^+)]
\\&=-16i\frac{[24]^4\langle13\rangle^2\langle23\rangle^2}{s_{12}s_{23}^2}-
16i\frac{[24]^4\langle13\rangle^2\langle23\rangle^2}{s_{12}s_{13}s_{23}^2}
\\&\sim\frac{[24]^4\langle23\rangle^2\langle13\rangle^2}{s_{12}s_{23}s_{13}}.
\end{split}
\end{equation}

So far, we have given all the independent three- and four-point
amplitudes, other three- and four-point amplitudes can be derived
from these amplitudes by using a parity transformation or performing
an appropriate replacement on the external legs. These results are
same with those given by KLT relations\cite{17}. Then the three- and
four-point tree amplitudes for gluons minimally coupled to graviton
satisfy $D_2$ relations. We then expect the $D_2$ relations hold in
all amplitudes where gluons are minimally coupled to gravitons.

\section{General discussions on $D_2$ relations for tree amplitudes in minimal coupling theory of gauge field and gravity}\label{general discussions}
In this section, we study general tree amplitudes in minimal
coupling theory of gauge field and gravity. $D_2$ relations can give
amplitudes with $N$ gluons and $M$ gravitons by sum of pure-gluon
partial amplitudes with $N+2M$ legs except for appropriate factors.
If all the gluons and gravitons take positive helicity, the
amplitude must vanish. This is because the pure-gluon amplitudes
with all legs of positive helicity vanish.

The tree amplitudes with one gluon of negative helicity and other
legs of positive helicity can be expressed as sum of pure-gluon
amplitudes with one leg of negative helicity and other legs of
positive helicity. Then these amplitudes also vanish. The amplitudes
with one graviton of negative helicity and other $N+M-1$ legs of
positive helicity are expressed by sum of MHV $(N+2M)$-gluon tree
amplitudes where the two gluons corresponding to the negative
helicity graviton take negative helicity. The two negative helicity
gluons in the $(N+2M)$-gluon amplitudes take the same momentum.
Considering the antisymmetry of the spinor products
\eqref{antisymmetry}, these MHV tree amplitudes for $N+2M$ gluons
vanish. Thus the amplitudes with one graviton of negative helicity
and other $N+M-1$ legs of positive helicity must vanish.

The results above given by $D_2$ relations are same with those given
by KLT relations. Then the $D_2$ relations can give the amplitudes
with all the legs of positive helicity for gauge-graviton minimal
coupling. They also give amplitudes with one leg of negative
helicity and other legs of positive helicity for gauge-gravity
minimal coupling.

Though the trivial cases are easy to consider, $D_2$ relations for
nontrivial helicity configurations are not so clear. In the
Subsections \ref{N gluons 1 graviton} and \ref{N gluons 2
gravitons}, we will give discussions on amplitudes with one and two
gravitons in addition to $N$ gluons. Here two gluons take negative
helicity and other legs take positive helicity. In Subsection \ref{N
gluons M gravitons}, we extend the discussions to arbitrary tree
amplitudes where $N$ gluons are minimally coupled to $M$ gravitons.
Two gluons take negative helicity and the other $N+M-2$ legs take
positive helicity. We then extend the $D_2$ relations for amplitudes
with one graviton in addition to $N$ gluons to relations independent
of the helicity configurations. We suggest $D_2$ relations should
hold for any helicity configurations. The $D_2$ relation for a given
amplitude is not unique. Different relations can be related by
relations among partial amplitudes of gauge field.

\subsection{$D_2$ relations for tree amplitudes with one positive helicity
graviton,
$N-2$ positive helicity gluons and two negative helicity
gluons}\label{N gluons 1 graviton}

\begin{figure}[tbp]
\begin{center}
\includegraphics[width=1\textwidth]{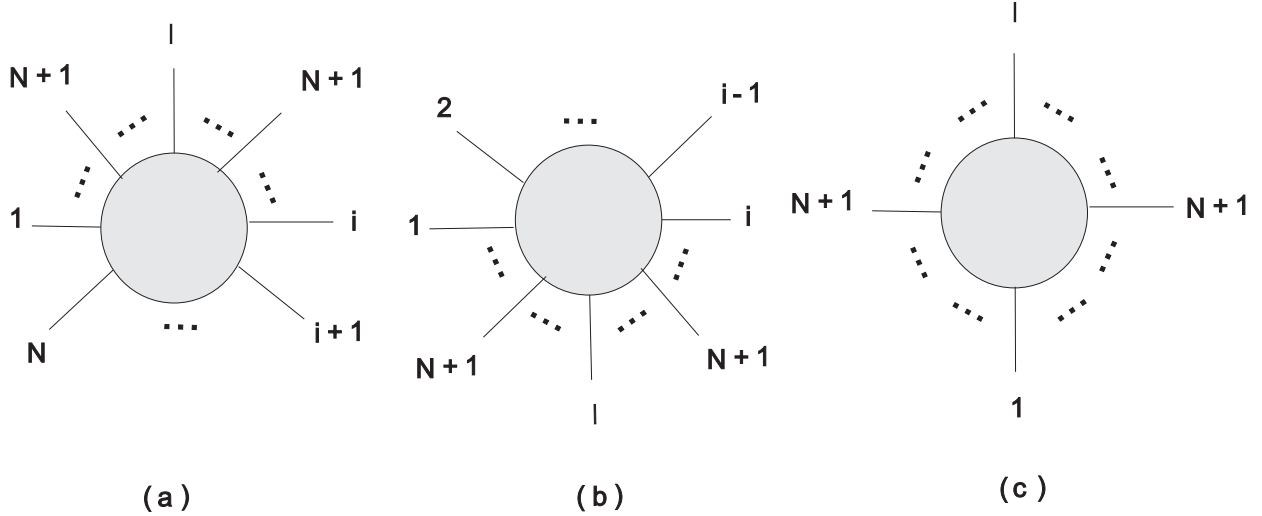}
\end{center}
\caption{Positions of the two gluons corresponding to the graviton
for (a) $1<l<i$, (b)  $i<l\leq N$ and (c) the expression independent
of helicity configuration. } \label{fig1}
\end{figure}

 The tree amplitudes with two gluons of negative helicity and
other legs of positive helicity can be expressed by\cite{28, 29}
\begin{equation}\label{MHV NgMh}
\begin{split}
&\mathcal {A}(1_g^-, 2_g^+, ..., i_g^-, ..., N_g^+, (N+1)_h^+, ...,
(N+M)_h^+)
\\&=ig^{N-2}\left(-\frac{\kappa}{2}\right)^M\frac{\langle1
i\rangle^4}{\langle12\rangle\langle23\rangle...\langle
N1\rangle}S(1, i, \{h^+\},\{g^+\}),
\end{split}
\end{equation}
where
\begin{equation}\label{S}
\begin{split}
&S(i, j, \{h^+\},
\{g^+\})=\left(\prod\limits_{m\in\{h^+\}}\frac{d}{da_m}\right)
\\&\times\prod\limits_{l\in\{g^+\}}\exp{\left[\sum\limits_{n_1\in\{h^+\}}a_{n_1}\frac{\langle
li\rangle\langle lj\rangle[l n_1]}{\langle n_1 i\rangle\langle n_1
j\rangle\langle ln_1\rangle}
\times\exp{\left[\sum\limits_{n_2\in\{h^+\},n_2\neq
n_1}a_{n_2}\frac{\langle n_1i\rangle\langle n_1 j\rangle[n_1
n_2]}{\langle n_2 i\rangle\langle n_2 j\rangle\langle n_1
n_2\rangle}\exp{\left[...\right]}\right] }\right]}\Bigg|_{a_j=0}.
\end{split}
\end{equation}

In $M=1$ case where all the legs take positive helicity except $1$
and $i$, the amplitude is reduced to
\begin{equation}\label{Ng1h_1}
\begin{split}
&\mathcal {A}(1_g^-, 2_g^+, ..., i_g^-, ..., N_g^+,
(N+1)_h^+)
\\=&ig^{N-2}\left(-\frac{\kappa}{2}\right)\frac{\langle1i\rangle^4}{\langle12\rangle\langle23\rangle...\langle
N1\rangle}\sum\limits_{l\in\{g^+\}}\frac{\langle l1\rangle\langle li
\rangle[l, N+1]}{\langle N+1, 1\rangle\langle N+1, i\rangle\langle
l, N+1 \rangle}
\\=&ig^{N-2}\left(\frac{\kappa}{2}\right)\sum\limits_{l\in\{g^+\}}\frac{\langle1i\rangle^4}{\langle12\rangle\langle23\rangle...\langle
N1\rangle}\langle l, N+1\rangle[l, N+1]
\\\times&\frac{\langle1l\rangle}{\langle1,N+1\rangle\langle N+1,
l\rangle}\frac{\langle li\rangle}{\langle l,N+1\rangle\langle
N+1,i\rangle}.
\end{split}
\end{equation}
In the equation above,
$\frac{\langle1i\rangle^4}{\langle12\rangle\langle23\rangle...\langle
N1\rangle}$ is just the MHV amplitude for $N$ gluons with the
permutation $1,...,N$. Using \eqref{property1}, we have $\langle l,
N+1\rangle[l, N+1]=s_{l, N+1}$. With the eikonal
identity\eqref{eikonal identity},
$\frac{\langle1l\rangle}{\langle1,N+1\rangle\langle N+1, l\rangle}$
and $\frac{\langle li\rangle}{\langle l,N+1\rangle\langle
N+1,i\rangle}$ can split into sums of terms over all the legs
between $1$, $l$ and $l$, $i$ respectively. For $1<l<i$,
\begin{equation}
\begin{split}\label{split1}
\frac{\langle1l\rangle}{\langle1,N+1\rangle\langle
N+1,l\rangle}&=\sum\limits_{r=1}^{l-1}\frac{\langle
r,r+1\rangle}{\langle r,N+1\rangle\langle N+1,r+1\rangle},
\\\frac{\langle li\rangle}{\langle l,N+1\rangle\langle
N+1,i\rangle}&=\sum\limits_{t=l}^{i-1}\frac{\langle
t,t+1\rangle}{\langle t,N+1\rangle\langle N+1,t+1\rangle}.
\end{split}
\end{equation}
For $i<l\leq N$,
\begin{equation}
\begin{split}\label{split2}
\frac{\langle il\rangle}{\langle i,N+1\rangle\langle
N+1,l\rangle}&=\sum\limits_{r=i}^{l-1}\frac{\langle
r,r+1\rangle}{\langle r,N+1\rangle\langle N+1,r+1\rangle},
\\\frac{\langle l1\rangle}{\langle l,N+1\rangle\langle
N+1,1\rangle}&=\sum\limits_{t=l}^{N}\frac{\langle
t,t+1\rangle}{\langle t,N+1\rangle\langle N+1,t+1\rangle},
\end{split}
\end{equation}
where we define $t+1=1$ for $t=N$. The amplitude then becomes
\begin{equation}\label{Ng1h}
\begin{split}
&\mathcal {A}(1_g^-, 2_g^+, ..., i_g^-, ..., N_g^+, (N+1)_h^+)
\\&=ig^{N-2}\left(\frac{\kappa}{2}\right)\Bigg(\sum\limits_{1<l<i}s_{l,
N+1}\sum\limits_{r=1}^{l-1}\sum\limits_{t=l}^{i-1}+\sum\limits_{i<l\leq
N}s_{l, N+1}\sum\limits_{r=i}^{l-1}\sum\limits_{t=l}^{N}\Bigg)
\\&\cdot\frac{\langle1i\rangle^4}{\langle12\rangle\langle23\rangle...\langle
N1\rangle}\frac{\langle r,r+1\rangle}{\langle r,N+1\rangle\langle
N+1,r+1\rangle}\frac{\langle t,t+1\rangle}{\langle
t,N+1\rangle\langle N+1,t+1\rangle},
\end{split}
\end{equation}
where for $t=N$ in the sum over $t$, we define $k_{t+1}\equiv k_1$.
For a given $r$, the numerator of $\frac{\langle
r,r+1\rangle}{\langle r,N+1\rangle\langle N+1,r+1\rangle}$ is a
spinor product for two adjacent points. Then we can remove $\langle
r,r+1\rangle$ in $\frac{\langle r,r+1\rangle}{\langle
r,N+1\rangle\langle N+1,r+1\rangle}$ and
$\frac{\langle1i\rangle^4}{\langle12\rangle\langle23\rangle...\langle
N1\rangle}$ simultaneously. Then $\langle r,r+1\rangle$ in the
denominator of
$\frac{\langle1i\rangle^4}{\langle12\rangle\langle23\rangle...\langle
N1\rangle}$ is replaced by  $\langle r,N+1\rangle\langle
N+1,r+1\rangle$. After this replacement,
$\frac{\langle1i\rangle^4}{\langle12\rangle\langle23\rangle...\langle
N1\rangle}$ becomes
$\frac{\langle1i\rangle^4}{\langle12\rangle\langle23\rangle...\langle
r,N+1\rangle\langle N+1,r+1\rangle...\langle N1\rangle}$ which is
the ($N+1$)-gluon MHV tree amplitude with the permutation $1$,
$2$,...,$r$, $N+1$, $r+1$,...,$N$. Thus $\frac{\langle
r,r+1\rangle}{\langle r,N+1\rangle\langle N+1,r+1\rangle}$ just
insert a gluon with momentum $k_{N+1}$ between the two gluons $r$
and $r+1$. For a given $l$ ($1<l<i$), the sum over $r$ for $1\leq
r\leq l-1$ becomes sum over all the possible insertions between $1$
and $l$. For $i<l\leq N$, the sum over $r$ for $i\leq r\leq l-1$
becomes sum over all the possible insertions between $i$ and $l$.
After a similar discussion, $\frac{\langle t,t+1\rangle}{\langle
t,N+1\rangle\langle N+1,t+1\rangle}$ insert a gluon with momentum
$k_{N+1}$ at the positions between $l$, $i$ for $1<l<i$ and between
$i$, $1$ for $i<l\leq N$. The two gluons with momentum $k_{N+1}$
just correspond to the left- and right-moving sectors of the
graviton\footnote{In this section, we let the gluons corresponding
to a graviton take the momentum of the graviton for convenience.
This is a little different from in the sections above where each
gluon from a graviton take half of the momentum of the graviton.
However, a redefinition of the momentum $k\rightarrow\frac{1}{2}k$
only contribute a constant factor which
 does not affect our discussions. }. Then
$\frac{\langle1i\rangle^4}{\langle12\rangle\langle23\rangle...\langle
N1\rangle}$ becomes
$\frac{\langle1i\rangle^4}{\langle12\rangle\langle23\rangle...\langle
r,N+1\rangle\langle N+1,r+1\rangle...\langle t,N+1\rangle\langle
N+1,t\rangle...\langle N1\rangle}$ which is the MHV tree amplitude
with $N+2$ gluons.

Thus, the amplitude can be considered as a sum of terms. In each
term, there is an MHV tree amplitude for $N+2$ gluons. Two of the
$N+2$ gluons take the momentum $k_{N+1}$. Then the amplitude
satisfies the relation
\begin{equation}\label{MHV Ng1h}
\begin{split}
&\mathcal {A}(1_g^-, 2_g^+, ..., i_g^-, ..., N_g^+, (N+1)_h^+)
\\=&ig^{N-2}\left(\frac{\kappa}{2}\right)
\sum\limits_{l\in\{g^+\}}s_{l,N+1}\sum\limits_P\mathcal
{A}_{MHV}^{N+2}(P),
\end{split}
\end{equation}
Where for any given $l$ in $\{g^+\}$, we sum over permutations $P$.
$P$ are the permutations in which the relative position of the $N$
gluons is $1_g$, $2_g$,..., $N_g$, one gluon corresponding to the
graviton $(N+1)_h$ can be inserted at any position between $1_g$ and
$l_g$, the other gluon corresponding to the graviton can be inserted
at any position between $l_g$ and $i_g$ (See Fig. \ref{fig1}(a) and
(b)).

\subsection{$D_2$ relations for tree amplitudes with two positive helicity
gravitons,
$N-2$ positive helicity gluons and two negative helicity gluons
}\label{N gluons 2 gravitons}
 The tree amplitude \eqref{MHV
NgMh} and \eqref{S}, with $M=2$ can be reduced to
\begin{equation}
\mathcal {A}(1_g^-, 2_g^+, ..., i_g^-, ...,N_g^+, (N+1)_h^+,
(N+2)_h^+)=\mathbb{A}+\mathbb{B}+\mathbb{C},
\end{equation}
where $\mathbb{A}$, $\mathbb{B}$ and $\mathbb{C}$ are
\begin{align}
\label{A part}
\nonumber\mathbb{A}&=ig^{N-2}\left(-\frac{\kappa}{2}\right)^2\frac{\langle1i\rangle^4}{\langle12\rangle\langle23\rangle...\langle
N1\rangle}
\\&\times \sum\limits_{l\in\{g^+\}}s_{N+1, N+2}s_{l,N+1}\frac{\langle1l\rangle}{\langle1,N+1\rangle\langle
N+1,l\rangle}\frac{\langle li\rangle}{\langle l,N+1\rangle\langle
N+1,i\rangle}
\\\nonumber&\times\frac{\langle1,N+1\rangle}{\langle1,N+2\rangle\langle
N+2,N+1\rangle}\frac{\langle N+1,i\rangle}{\langle
N+1,N+2\rangle\langle N+2,i\rangle},
\end{align}
\begin{align}
\label{B
part}\nonumber\mathbb{B}&=ig^{N-2}\left(-\frac{\kappa}{2}\right)^2\frac{\langle1i\rangle^4}{\langle12\rangle\langle23\rangle...\langle
N1\rangle}
\\&\times \sum\limits_{l\in\{g^+\}}s_{N+2, N+1}s_{l,N+2}\frac{\langle1l\rangle}{\langle1,N+2\rangle\langle
N+2,l\rangle}\frac{\langle li\rangle}{\langle l,N+2\rangle\langle
N+2,i\rangle}
\\\nonumber&\times\frac{\langle1,N+2\rangle}{\langle1,N+1\rangle\langle
N+1,N+2\rangle}\frac{\langle N+2,i\rangle}{\langle
N+2,N+1\rangle\langle N+1,i\rangle},
\end{align}
\begin{align}
\label{C
part}\nonumber\mathbb{C}&=ig^{N-2}\left(-\frac{\kappa}{2}\right)^2\frac{\langle1i\rangle^4}{\langle12\rangle\langle23\rangle...\langle
N1\rangle}
\\&\times\sum\limits_{l\in\{g^+\}}s_{l,N+2}\frac{\langle l1\rangle}{\langle N+2,1\rangle\langle l,N+2\rangle}
\frac{\langle li\rangle}{\langle l,N+2\rangle\langle N+2,i\rangle}
\\\nonumber&\times\sum\limits_{k\in\{g^+\}}s_{k,N+1}\frac{\langle k1\rangle}{\langle N+1,1\rangle\langle l,N+1\rangle}
\frac{\langle ki\rangle}{\langle k,N+1\rangle\langle N+1,i\rangle}.
\end{align}

\begin{figure}[tbp]
\begin{center}
\includegraphics[width=0.7\textwidth]{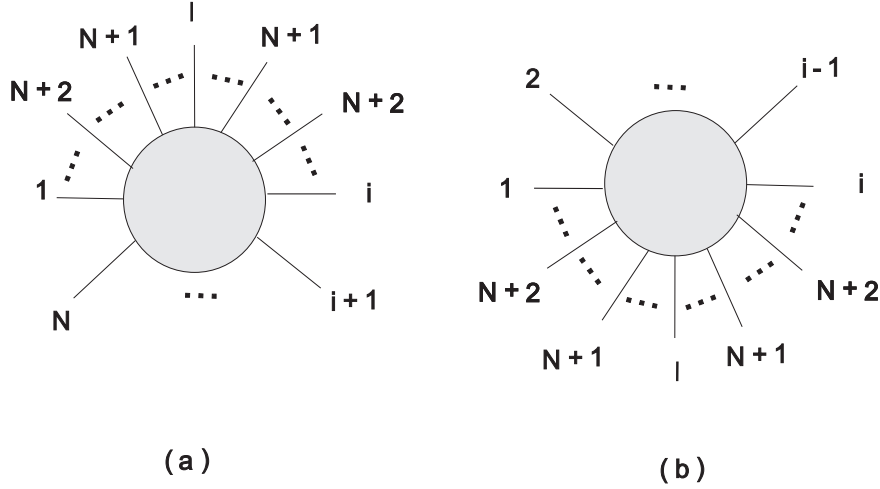}
\end{center}
\caption{Positions of the gluons corresponding to the two gravitons
$(N+1)_h$ and $(N+2)_h$ for (a) $1<l<i$, (b)  $i<l\leq N$ in
$\mathbb{A}$ part. }\label{fig2}
\end{figure}
We first  look at $\mathbb{A}$ part.
$\frac{\langle1l\rangle}{\langle1,N+1\rangle\langle N+1,l\rangle}$
and $\frac{\langle li\rangle}{\langle l,N+1\rangle\langle
N+1,i\rangle}$ can split into sums of terms as in \eqref{split1} and
\eqref{split2}. For a given $l$, as in $M=1$ case, they insert the
two gluons corresponding to the graviton $(N+1)_h$ into positions
between $1$, $l$ and $l$, $i$ respectively. After this insertion, we
consider the insertion of gluons corresponding to $(N+2)_h$. With
the eikonal identity\eqref{eikonal identity}, for $1<l<i$,
$\frac{\langle1,N+1\rangle}{\langle1,N+2\rangle\langle
N+2,N+1\rangle}$ and $\frac{\langle N+1,i\rangle}{\langle
N+1,N+2\rangle\langle N+2,i\rangle}$ in \eqref{A part} can be given
as
\begin{equation}\label{split3}\begin{split}
\frac{\langle1,N+1\rangle}{\langle1,N+2\rangle\langle
N+2,N+1\rangle}&=\left(\frac{\langle1,r\rangle}{\langle1,N+2\rangle\langle
N+2,r\rangle}+\frac{\langle r,N+1\rangle}{\langle
r,N+2\rangle\langle N+2,N+1\rangle}\right),
\\\frac{\langle N+1,i\rangle}{\langle
N+1,N+2\rangle\langle N+2,i\rangle}&=\left(\frac{\langle
t+1,i\rangle}{\langle t+1,N+2\rangle\langle
N+2,i\rangle}+\frac{\langle N+1,t+1\rangle}{\langle
N+1,N+2\rangle\langle N+2,t+1\rangle}\right),
\end{split}
\end{equation}
while for $i<l\leq N$,
\begin{equation}\label{split4}\begin{split}
\frac{\langle i,N+1\rangle}{\langle i,N+2\rangle\langle
N+2,N+1\rangle}&=\left(\frac{\langle i,r\rangle}{\langle
i,N+2\rangle\langle N+2,r\rangle}+\frac{\langle
r,N+1\rangle}{\langle r,N+2\rangle\langle N+2,N+1\rangle}\right),
\\\frac{\langle N+1,1\rangle}{\langle
N+1,N+2\rangle\langle N+2,1\rangle}&=\left(\frac{\langle
t+1,1\rangle}{\langle t+1,N+2\rangle\langle
N+2,1\rangle}+\frac{\langle N+1,t+1\rangle}{\langle
N+1,N+2\rangle\langle N+2,t+1\rangle}\right).
\end{split}
\end{equation}
The first term of the sum in each line of \eqref{split3} and
\eqref{split4} can split into sum over adjacent points again, for
example, the first term in the first line of \eqref{split3} can be
expressed as
\begin{equation}\label{split3.1}
\sum\limits_{p=1}^{r-1}\frac{\langle p,p+1\rangle}{\langle
p,N+2\rangle\langle N+2,p+1\rangle}.
\end{equation}
For a given $r$, we have inserted a gluon corresponding to $(N+1)_h$
between $r$ and $r+1$, then \eqref{split3.1} insert a gluon
corresponding to $(N+2)_h$ at a position between $1$ and $r$. The
second term in the first line of \eqref{split3} insert an gluon
corresponding to $(N+2)_h$ between $r$ and the gluon corresponding
to $(N+1)_h$. Thus the first line of \eqref{split3} just insert a
gluon corresponding to $(N+2)_h$ at the positions between $1$ and
the gluon corresponding to $(N+1)_h$, where the gluon corresponding
to $(N+1)_h$ have been inserted at positions between $1$ and $l$. In
a same way, the second line of \eqref{split3} insert the other gluon
corresponding to $(N+2)_h$ at positions between the gluon
corresponding to $(N+1)_h$ and $i$, where this gluon corresponding
to $(N+1)_h$ have been inserted at a position between $l$ and
$i$(See Fig. \ref{fig2} (a)). Following a similar discussion, for
the case of $i<l\leq N$, we insert the two gluons corresponding to
$(N+1)_h$ at positions between $i$, $l$ and $l$, $1$ respectively.
Then insert one gluon corresponding to $(N+2)_h$ at the positions
between $i$ and the gluon corresponding to $(N+1)_h$ which is
between $i$, $l$. Insert the other gluon corresponding to $(N+2)_h$
at the positions between the other gluon corresponding to $(N+1)_h$
and $1$(See Fig. \ref{fig2} (b)).  The $\mathbb{A}$ part then
satisfy the relation
\begin{equation}
\mathbb{A}=\sum\limits_{l\in\{g^+\}}s_{l,N+1}s_{N+1,N+2}\sum_{P_1}\mathcal
{A}_{MHV}^{N+4}(P_1),
\end{equation}
where for a  given $l$, $P_1$ are the possible insertions of the
gluons corresponding to the two gravitons. These insertions has the
form $1$, ..., $N+2$, ..., $N+1$,...,$l$, ..., $N+1$, ..., $N+2$,
..., $i$, ..., $N$ for $1<l<i$ and $1$,..., $i$, ..., $N+2$, ...,
$N+1$, ..., $l$, ..., $N+1$, ..., $N+2$, ... for $i<l\leq N$.

\begin{figure}[tbp]
\begin{center}
\includegraphics[width=0.7\textwidth]{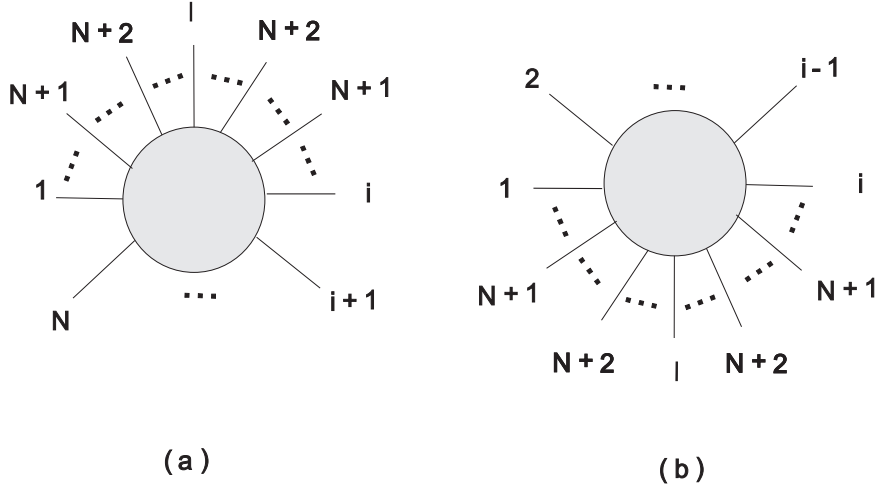}
\end{center}
\caption{Positions of the gluons corresponding to the two gravitons
$(N+1)_h$ and $(N+2)_h$  for (a) $1<l<i$, (b) $i<l\leq N$ in
$\mathbb{B}$ part. }\label{fig3}
\end{figure}

The $\mathbb{B}$ part can be derived from $\mathbb{A}$ part by the
replacement $N+1\leftrightarrow N+2$ and is given as
\begin{equation}
\mathbb{B}=\sum\limits_{l\in\{g^+\}}s_{l,N+2}s_{N+2,N+1}\sum_{P_2}\mathcal
{A}_{MHV}^{N+4}(P_2),
\end{equation}
where for a  given $l$, $P_2$ are the possible insertions of the
gluons corresponding to the two gravitons. These insertions has the
form $1$, ..., $N+1$, ..., $N+2$,...,$l$, ..., $N+2$, ..., $N+1$,
..., $i$, ..., $N$ for $1<l<i$( See Fig. \ref{fig3} (a)) and
$1$,..., $i$, ..., $N+1$, ..., $N+2$, ..., $l$, ..., $N+2$, ...,
$N+1$, ... for $i<l\leq N$(See Fig. \ref{fig3} (b)).

\begin{figure}[tbp]
\begin{center}
\includegraphics[width=0.7\textwidth]{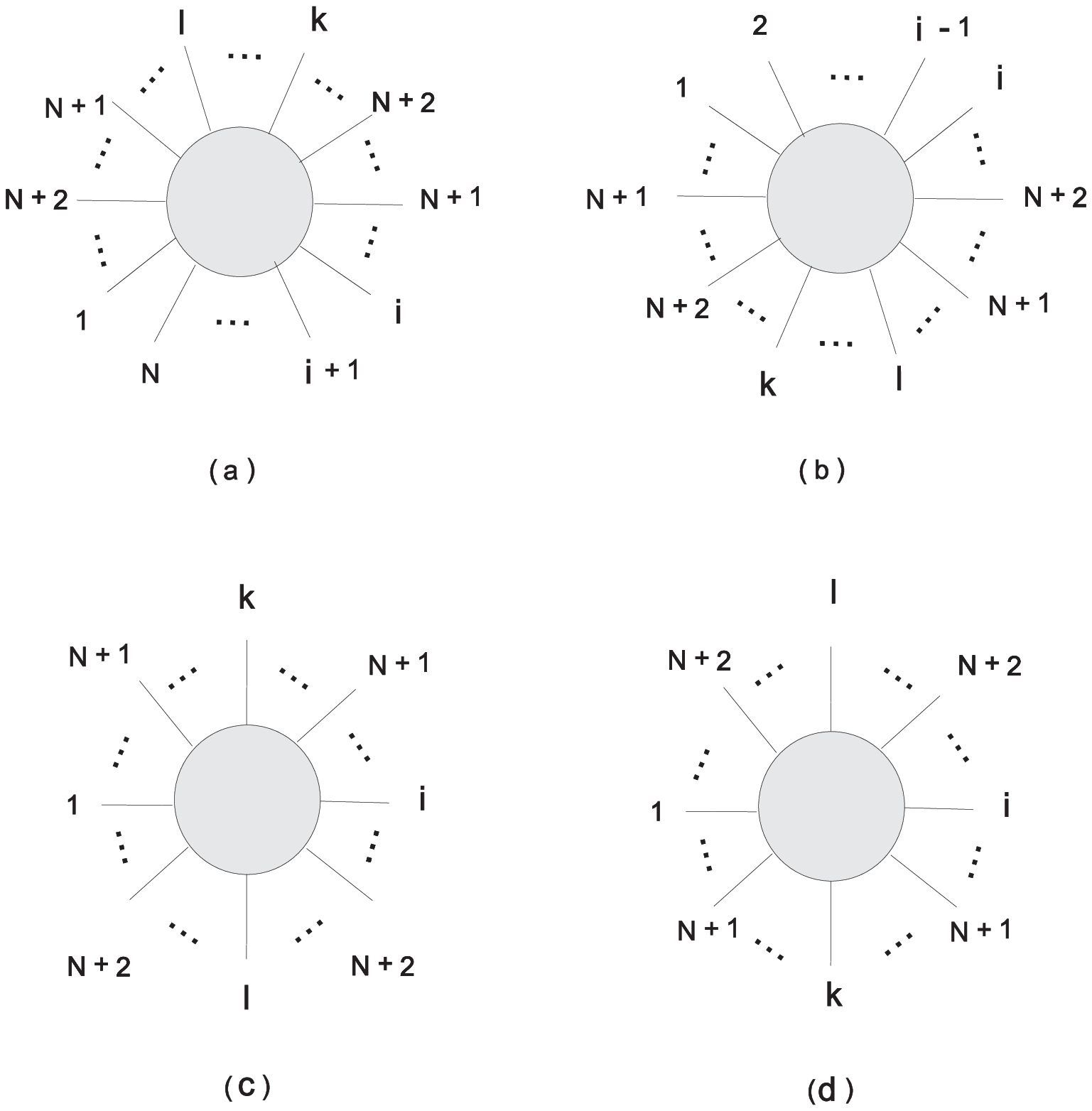}
\end{center}
\caption{Positions of the gluons corresponding to the two gravitons
$(N+1)_h$ and $(N+2)_h$  for (a) $1<l<i$, $1<k<i$, (b) $i<l\leq N$,
$i<k\leq N$ in (c) $i<l\leq N$, $1<k<i$ and (d) $1<l<i$, $i<k\leq N$
in $\mathbb{C}$ part. Here we first insert the two gluons
corresponding to $(N+2)_h$ between $1$, $l$ and $l$, $i$
respectively. We then insert two gluons corresponding to $(N+1)_h$
between $1$, $k$ and $k$, $i$ respectively. }\label{fig4}
\end{figure}

Now we consider the $\mathbb{C}$ part. Both the second and the third
lines in \eqref{C part}, can split into the forms of \eqref{split1}
and \eqref{split2}. Then for a given $l$, each term in the second
line of the expression \eqref{C part} insert the two gluons
corresponding to $(N+2)_h$ at the positions between $1$, $l$ and
$l$, $i$ respectively. Then for any given $k$, each term in the
second line of \eqref{C part} insert the two gluons corresponding to
$(N+1)_h$ at the positions between $1$, $k$ and $k$, $i$
respectively. The $\mathbb{C}$ part then becomes
\begin{equation}
\mathbb{C}=\sum\limits_{l,k\in{g^+}}s_{l,N+2}s_{k,N+1}\sum\limits_{P_3}\mathcal
{A}_{MHV}^{N+4}(P_3).
\end{equation}
where $P_3$ are the insertions of the four gluons corresponding to
the two gravitons $(N+2)_h$ and $(N+1)_h$. In these insertions, two
gluons corresponding to $(N+2)_h$ are inserted at the positions
between $1$, $l$ and $l$, $i$ respectively, then two gluons
corresponding to $(N+1)_h$ are inserted at the positions between
$1$, $k$ and $k$, $i$ respectively( See Fig. \ref{fig4}).

After considering all the contributions from $\mathbb{A}$,
$\mathbb{B}$ and $\mathbb{C}$, we give the $D_2$ relation
\begin{equation}
\begin{split} &\mathcal {A}(1_g^-, 2_g^+, ..., i_g^-, ...,N_g^+, (N+1)_h^+,
(N+2)_h^+)
\\=&\sum\limits_{l\in\{g^+\}}s_{l,N+1}s_{N+1,N+2}\sum_{P_1}\mathcal
{A}_{MHV}^{N+4}(P_1)+\sum\limits_{l\in\{g^+\}}s_{l,N+2}s_{N+2,N+1}\sum_{P_2}\mathcal
{A}_{MHV}^{N+4}(P_2)
\\+&\sum\limits_{l,k\in{g^+}}s_{l,N+2}s_{k,N+1}\sum\limits_{P_3}\mathcal
{A}_{MHV}^{N+4}(P_3).
\end{split}
\end{equation}

\subsection{$D_2$ relations
for arbitrary tree amplitudes with $N$ gluons minimally coupled to
$M$ gravitons}\label{N gluons M gravitons} The amplitudes with more
gravitons are more complicated. However the discussions are similar
with those for amplitudes with one and two gravitons. The amplitude
 with $N$ gluons minimally coupled to
$M$ gravitons, where two gluons take negative helicity and other
legs take positive helicity \eqref{MHV NgMh} and \eqref{S} can be
given by a sum of terms. Each term has the form
\begin{equation}
\begin{split}
&ig^{N-2}\left(-\frac{\kappa}{2}\right)^{M}\frac{\langle1i\rangle^4}{\langle12\rangle\langle23\rangle...\langle
N1\rangle}
\\\times&\frac{\langle l_11\rangle\langle l_1i\rangle[l_1n_1^1]}{\langle n_1^11\rangle\langle n_1^1i\rangle\langle n_1^1n_2^1\rangle}
\times\frac{\langle n_1^11\rangle\langle n_1^1i\rangle
[n_1^1n_2^1]}{\langle n_2^11\rangle\langle n_2^1i\rangle\langle
n_1^1n_2^1\rangle}\times...
\\\times&\frac{\langle l_21\rangle\langle l_2i\rangle[l_2n_1^2]}{\langle n_1^21\rangle\langle n_1^2i\rangle\langle n_1^2n_2^2\rangle}
\times\frac{\langle n_1^21\rangle\langle n_1^2i\rangle
[n_1^2n_2^2]}{\langle n_2^21\rangle\langle n_2^2i\rangle\langle
n_1^2n_2^2\rangle}\times...
\\\vdots
\\\times&\frac{\langle l_N1\rangle\langle l_Ni\rangle[l_Nn_1^N]}{\langle n_1^N1\rangle\langle n_1^Ni\rangle\langle n_1^Nn_2^N\rangle}
\times\frac{\langle n_1^N1\rangle\langle n_1^Ni\rangle
[n_1^Nn_2^N]}{\langle n_2^N1\rangle\langle n_2^Ni\rangle\langle
n_1^Nn_2^N\rangle}\times...,
\end{split}
\end{equation}
where each $l_i$ can be any positive helicity gluon, and $n_1^1$,
$n_2^1$,..., $n_1^2$, $n_2^2$,..., ..., $n_1^N$, $n_2^N$,... is  a
permutation of all the gravitons. Following the discussions on
amplitudes with one and two gravitons, this term can be given by sum
of MHV amplitudes with $N+2M$ gluons with appropriate factors. The
first line insert two gluons corresponding to $n_1^1$ between $1$,
$l_1$ and $l_1$, $i$ respectively, then insert two gluons
corresponding to $n_2^1$ between $1$, one gluon corresponding to
$n_1^1$ and $n_1^1$, the other gluon corresponding to $n_1^1$
respectlvely, ...  After inserting all the gluons corresponding to
gravitons in the first line, we insert two gluons corresponding to
$n_1^2$ between $1$, $l_2$ and $l_2$, $i$ respectively, then insert
two gluons corresponding to $n_2^2$ between $1$, one gluon
corresponding to $n_1^2$ and $n_1^2$, the other gluon corresponding
to $n_1^2$ respectively, ... In this way, we insert all the $2M$
gluons corresponding to the $M$ gravitons into the amplitudes. The
phase factor is $s_{l_1, n_1^1}s_{n_1^1, n_2^1}...s_{l_2,
n_1^2}s_{n_1^2, n_2^2}...s_{l_N, n_1^N}s_{l_N, n_2^N}...$. At last,
the amplitudes become MHV tree amplitudes with $N+2M$ gluons, where
the two gluons corresponding to a same graviton take the same
momentum. This is just the $D_2$ relations in field theory.

In string theory, the $D_2$ relations are independent of helicity
configurations of the legs. Then we expect the $D_2$ relations
should have helicity-independent form. For example, in the relation
for amplitudes with one graviton and $N$ gluons given in Subsection
\ref{N gluons 1 graviton}, we only sum over $l$ correspongding to
the gluons with positive helicity, and for each $l$, the two gluons
are inserted at the positions between $1$, $l$ and $l$, $i$
respectively, the relation \eqref{MHV Ng1h} depends on the relative
positions  of the two negative helicity gluons.  Then we expect the
relation \eqref{MHV Ng1h} can be extended to that independent of the
relative positions of the two negative helicity legs. In fact, in
the expression \eqref{Ng1h_1}, we can sum over $l$  ($1< l\leq N$).
This is because $\langle li\rangle$ vanishes for $l=i$. Using the
eikonal identity \eqref{eikonal identity} and the identity
\eqref{momentum conservation} implied by momentum conservation, the
amplitude becomes
\begin{equation}
\begin{split}
&\mathcal {A}(1_g^-, 2_g^+, ..., i_g^-, ..., N_g^+, (N+1)_h^+)
\\=&ig^{N-2}\left(\frac{\kappa}{2}\right)\sum\limits_{1< l\leq
N}\frac{\langle1i\rangle^4}{\langle12\rangle\langle23\rangle...\langle
N1\rangle}\langle l, N+1\rangle[l, N+1]
\\\times&\frac{\langle1l\rangle}{\langle1,N+1\rangle\langle N+1,
l\rangle}\frac{\langle l1\rangle}{\langle l,N+1\rangle\langle
N+1,1\rangle}.
\end{split}
\end{equation}
Then we repeat the discussions above. The amplitude satisfy the
relation
\begin{equation}
\begin{split}
&\mathcal {A}(1_g^-, 2_g^+,  ...,i_g^-,...,  N_g^+, (N+1)_h^+)
\\=&ig^{N-2}\left(\frac{\kappa}{2}\right)
\sum\limits_{1< l\leq N}s_{l,N+1}\sum\limits_{P'}\mathcal
{A}_{MHV}^{N+2}(P'),
\end{split}
\end{equation}
where we sum over all the external gluons. For any given $l$, $P'$
in this relation denote all the permutations where one of the two
gluons corresponding to the graviton is inserted at positions to the
left of $l$ and right of $1$, the other one is inserted to the left
of $1$ and right of $l$(See Fig. \ref{fig1}(c)). Since this
expression of the amplitude does not depend on the helicity
configuration of the legs.  We suggest that for any helicity
configuration, the tree amplitude with $N$ gluons minimally coupled
to one graviton satisfy the relation
\begin{equation}\label{general Ng1h}
\begin{split}
&\mathcal {A}(1_g, 2_g, ..., N_g, (N+1)_h)
\\=&ig^{N-2}\left(\frac{\kappa}{2}\right)
\sum\limits_{1< l\leq N}s_{l,N+1}\sum\limits_{P'}\mathcal
{A}^{N+2}(P').
\end{split}
\end{equation}
Though this extension will be more complicated, we expect there must
be such extensions to the relations for arbitrary helicity
configurations.

The $D_2$ relation for a given amplitude may have different
expressions as in string theory. A tree amplitude for gauge-gravity
minimal coupling can be expressed by different sets of pure-gluon
partial tree amplitudes. The permutations of the legs and the
factors in different expressions are different. However, the partial
tree amplitudes of gluons are not independent of each other, there
are relations among the partial tree amplitudes with gluons\cite{6,
30}. Then the different expressions of the $D_2$ relation for a
given amplitude can be related by the relations among pure-gluon
partial tree amplitudes. To see this, we take amplitude with three
gluons and one graviton as an example. In Section \ref{KLT vs disk},
the relation is given by \eqref{3g1h}, the factor is in
$s_{13}$-channel. However, in Section \ref{general discussions}, the
relation is given by \eqref{general Ng1h}. For $N=3$, we have
\begin{equation}\label{3g1h_1}
\begin{split}
\mathcal{A}(1_g, 2_g, 3_g, 4_h)&= g\left(-\frac{\kappa}{2}
\right)\Bigg[s_{24}\mathcal {A}(1_g, 5_g, 2_g, 4_g,
3_g)+s_{24}\mathcal {A}(1_g, 4_g, 2_g, 3_g, 5_g)
\\&+s_{34}\mathcal
{A}(1_g, 4_g, 2_g, 3_g, 5_g)+s_{34}\mathcal {A}(1_g, 2_g, 4_g, 3_g,
5_g)\Bigg],
\end{split}
\end{equation}
where we denote the two gluons corresponding to $4_h$ by $4_g$ and
$5_g$. Then the amplitude is given by different expressions. In the
last expression, $s_{2,4}=s_{1,3}$, $s_{24}+s_{34}=-s_{14}=-s_{23}$
and $s_{34}=s_{12}$. The second and the third terms can be given by
one term with the factor $-s_{23}$. Using the relations among
partial amplitudes, we have
\begin{equation}\begin{split}
\mathcal {A}(1_g, 4_g, 2_g, 3_g, 5_g)&=\frac{s_{13}}{s_{23}}\mathcal
{A}(1_g,5_g, 2_g, 4_g, 3_g),
\\\mathcal {A}(1_g, 2_g, 4_g, 3_g,
5_g)&=\frac{s_{13}}{s_{12}}\mathcal {A}(1_g,5_g, 2_g, 4_g, 3_g).
\end{split}
\end{equation}
Then the two relations \eqref{3g1h_1} and \eqref{3g1h} are
equivalent.

\section{Conclusion}\label{conclusion}
In this paper, we study the amplitudes where gluons are minimally
coupled with gravitons. We find the three- and four-point amplitudes
satisfy the field theory limits of $D_2$ relations in string theory.
The left- and right-moving sectors are connected into a single one.

 We give particular
forms of the relations for the amplitude $\mathcal {A}(1_g^-, 2_g^+,
..., i_g^-, ..., N_g^+, (N+1)_h^+)$, and $\mathcal {A}(1_g^-, 2_g^+,
..., i_g^-, ..., N_g^+, (N+1)_h^+), (N+2)_h^+$. We extend the
relation to arbitrary helicity configurations for $N+1$ case.  The
discussions can be extended to arbitrary legs with arbitrary
helicity configurations. The tree amplitude with $N$ gluons and $M$
gravitons can be expressed by sum of amplitudes for  $N+2M$ gluons
with appropriate factors. The relation for a given amplitude is not
unique, because there are relations among pure-gluon partial
amplitudes.

Though the $D_2$ relations and KLT factorization relations only hold
on $D_2$ and $S_2$ respectively in string theory, the field theory
limits of both two relations hold in in minimal coupling theory of
gauge and gravity. This is because we have two different methods to
incorporate gauge degree of freedom in string theory.

\section*{Acknowledgement}
We would like to thank  C. Cao, Y. Q. Wang and Y. Xiao for useful
discussions. The work is supported in part by the NNSF of China
Grant No. 90503009, No. 10775116, and 973 Program Grant No.
2005CB724508.

\appendix
\section{Spinor helicity formalism }\label{spinor helicity}
Here we given the useful properties of spinor helicity
formalism\cite{24, 25, 26} Positive and negative helicity spinor
\begin{equation}
|i^{\pm}\rangle \equiv|k^{\pm}_i\rangle\equiv u_\pm(k_i)=v_\mp(k_i),
\langle i^{\pm}|\equiv\langle k^{\pm}_i|\equiv
\bar{u}_{\pm}(k_i)=\bar{v}_{\pm}(k_i),
\end{equation}
where $u$ and $v$ are positive and negative energy solutions of
Dirac equation.
\begin{equation}
\begin{split}
&\langle ij\rangle\equiv\langle
i^{-}|j^{+}\rangle=\sqrt{|s_{ij}|}e^{i\phi_{ij}},
\\&[ij]\equiv\langle
i^{+}|j^{-}\rangle=\sqrt{|s_{ij}|}e^{-i(\phi_{ij}+\pi)}.
\end{split}
\end{equation}

 Momentum
\begin{equation}
\langle i^{\pm}|\gamma^{\mu}|i^{\pm}\rangle=2k_i.
\end{equation}

Polarization vector
\begin{equation}
\epsilon^{\pm}_{\mu}(k, q)=\pm\frac{\langle
q^{\pm}|\gamma_{\mu}|k^{\mp}\rangle}{\sqrt{2}\langle
q^{\mp}|k^{\pm}\rangle},
\end{equation}
where $q$ is reference momentum, reflecting the freedom of on-shell
gauge transformation, $k$ is the vector boson momentum.

Useful properties:

\begin{equation}\label{property1}
\langle ij\rangle[ji]=s_{ij}
\end{equation}

antisymmetry \begin{equation}\label{antisymmetry}\langle
ij\rangle=-\langle ji\rangle, [ij]=-[ji], \langle
ii\rangle=[ii]=0,\end{equation}

Fierz rearrangement
\begin{equation}
\langle i^{+}|\gamma^{\mu}|j^{+}\rangle\langle
k^+|\gamma_{\mu}|l^+\rangle=2[ik]\langle lj\rangle,
\end{equation}

charge conjugation
\begin{equation}
\langle i^{+}|\gamma^{\mu}|j^{+}\rangle=\langle
j^{-}|\gamma^{\mu}|i^{-}\rangle,
\end{equation}

eikonal identity
\begin{equation}\label{eikonal identity}
\sum\limits_{i=j}^{k-1}\frac{\langle i, i+1\rangle}{\langle
iq\rangle\langle q, i+1\rangle}=\frac{\langle jk\rangle}{\langle
jq\rangle\langle qk\rangle}.
\end{equation}

 in an $N$-point amplitude, momentum conservation imply
\begin{equation}\label{momentum conservation}
\sum\limits_{i=1,i\neq k}^{n}[ji]\langle ik\rangle=0.
\end{equation}
The amplitudes with all positive helicity gluons are zero. The
amplitudes for gluons with one negative helicity and others positive
helicity are zero. The amplitude for $N$ gluons with two negative
and $N-2$ positive(MHV) helicities can be given\cite{6}
\begin{equation}\label{MHV gluon amplitude}
\mathcal {A}(1^+,...,i^-,...,j^-,...,n^+)=i\frac{\langle
ij\rangle^4}{\langle12\rangle\langle23\rangle...\langle n1\rangle}.
\end{equation}

\end{document}